\documentclass[11pt,a4paper]{article}
\usepackage{jheppub}
\usepackage{subcaption,mathrsfs}
\usepackage[T1]{fontenc}
\usepackage{lmodern}
\newcommand{\N}{\mathcal N}
\newcommand{\E}{\mathcal E}
\newcommand{\D}{\mathcal D}
\newcommand{\tr}{\operatorname{tr}}
\newcommand{\sinc}{\operatorname{sinc}}
\newcommand{\arsinh}{\operatorname{arsinh}}
\newcommand{\ket}[1]{|#1\rangle}
\newcommand{\bra}[1]{\langle#1|}
\newcommand{\proj}[1]{\ket{#1}\bra{#1}}
\makeatletter
\renewcommand\section{\@startsection{section}{1}{\z@}{-3.5ex\@plus-1ex\@minus-.2ex}{2.3ex\@plus.2ex}{\normalfont\large}}
\makeatother

\title{Phase-controlled recoverable entanglement in accelerated detector registers}
\author{Zhiming Huang}
\affiliation{School of Economics and Management, Wuyi University, Jiangmen 529020, China}
\emailAdd{465609785@qq.com}
\abstract{
We study recoverable entanglement in a four-qubit cluster-state register whose two detector qubits couple to a scalar field. A local preparation phase leaves the initial reference--register entanglement and detector marginal unchanged, but controls protection under collective noise. At the optimal phase, a fixed decoder recovers a Bell pair after any channel preserving the singlet and triplet sectors. For uniformly accelerated detectors at finite separation, an exact trace-norm formula shows that the same phase is optimal at every time within the secular weak-coupling dynamics. Small separation produces a slow decay, whose rate and phase-dependent amplitude we derive. Matching the local response to an inertial thermal bath gives an analytic optimum time for the entanglement contrast. A detector measurement and conditional phase correction recover a two-qubit teleportation resource. Spatial smearing and an ordered weak-coupling then small-separation limit specify the microscopic scope of the lifetime prediction.
}
\keywords{Quantum Dissipative Systems, Thermal Field Theory, Space-Time Symmetries}
\begin{document}
\maketitle

\section{Introduction}
\label{sec:intro}
The Unruh effect relates the local response of a uniformly accelerated detector in the Minkowski vacuum to a thermal response~\cite{Unruh1976}. An entangled register also samples correlations between different positions. Its evolution therefore depends on the field's spatial correlations as well as on the way information is encoded in the register. A comparison based on local temperature alone does not specify the fate of this encoded entanglement.

The choice of initial resource can also affect what remains useful after relativistic evolution. For example, the teleportation fidelity in a Schwarzschild-field setting can increase or decrease with Hawking temperature, depending on the initial state~\cite{Wu2023}. To interpret such behavior, one must identify the entanglement being measured and the operations available for recovering it. Here we study entanglement between one reference qubit and a three-qubit register, and allow joint operations within the register.

Cluster states provide a concrete encoding for this question. They are resources for measurement-based quantum computation~\cite{Raussendorf2001}. Recent work with a four-qubit cluster state found that the negativity across a one-versus-three partition remains maximal when only one register qubit acts as an Unruh--DeWitt detector~\cite{Li2026}. We first identify why this encoding tolerates noise on either of two designated detector qubits. We then let both qubits couple to the field, so that the environment can act on the pair jointly.

Quantum erasure correction and noiseless subsystems describe how logical information can survive changes in the physical state~\cite{Knill1997,Knill2000}. Decoherence-free subspaces relate protection to symmetries of the environmental coupling~\cite{Lidar1998}. In the encoding considered here, discarding either detector qubit still leaves maximal reference--register entanglement. Protection under noise on both detectors instead depends on whether their singlet and triplet subspaces are preserved. These subspaces, called sectors below, classify states by their symmetry under detector exchange.

The field supplies a separate source of structure. Two uniformly accelerated detectors coupled to a massless scalar field have spatial spectra that differ from those of inertial detectors in a thermal bath at the same Unruh temperature~\cite{Hu2015}. Their local pointlike spectra agree. This makes the detector pair a useful setting for separating local thermal response from the effect of correlations between two positions.

A small departure from exact collective coupling introduces two relaxation scales. Fast relaxation establishes a long-lived intermediate state, and a weaker process eventually removes the remaining correlations. This separation is the basis of metastability in open quantum dynamics~\cite{Macieszczak2016}. In the present problem, the preparation phase determines the amount of entanglement in the intermediate state, while detector separation determines how quickly it is lost.

Our first main result identifies the phase that protects the encoded resource. We compare locally equivalent states with the same initial negativity and the same reduced detector state. At the optimal phase, a fixed decoder recovers a Bell pair after every channel that preserves the singlet and triplet sectors. For the detector generator, we also derive a reduced trace-norm formula and prove that this phase maximizes negativity at every time, including at finite separation. A detector measurement followed by a conditional phase correction gives a two-qubit resource whose entanglement measures the remaining distinguishability of the initial sectors.

Our second main result gives the lifetime and environmental contrast of this resource. A small transverse separation opens a weak antisymmetric transition channel. We derive the resulting slow decay rate and phase-dependent amplitude. Matching the local detector response between accelerated vacuum and an inertial thermal bath leaves a simple ratio of slow rates. The limiting decay laws then determine the observation time at which their entanglement contrast is largest.

The finite-separation lifetime is derived within secular weak-coupling dynamics~\cite{Davies1974,Kaplanek2020,Han2025}. Spatial smearing defines the microscopic coupling~\cite{Schlicht2004}, and section~\ref{sec:domain} specifies the symmetry and convergence assumptions connecting this model to the lifetime formula.

Section~\ref{sec:encoding} defines the encoding, sectors and fixed decoder. Section~\ref{sec:model} obtains the collective transition rates from the field spectra. Section~\ref{sec:exact} derives the population equations, exact negativity and recovered resource. Section~\ref{sec:slow} uses these equations to obtain the slow lifetime and the matched-temperature contrast. Section~\ref{sec:domain} specifies the common-coupling protection, the lifetime approximation and the sensitivity to additional noise.

\section{Encoding, sector protection and decoding}
\label{sec:encoding}
Let $A$ be a reference qubit and $B,C,D$ a register. Only $C,D$ interact with the field. The qubits $A,B$ are spectators, meaning that they have no detector--field coupling. We use the computational basis $\ket0,\ket1$ and order the qubits as $ABCD$. In the interaction picture, with $\hbar=c=k_B=1$, define the normalized negativity by
\begin{equation}
 \N(\rho)=\|\rho^{T_A}\|_1-1,\qquad
 \|Q\|_1=\tr\sqrt{Q^\dagger Q}.
 \label{eq:negativity}
\end{equation}

Here $T_A$ denotes partial transpose on the reference qubit: the two matrix indices of $A$ are exchanged while the register indices are unchanged. The symbol $\tr$ denotes the matrix trace. The operator $Q$ is an arbitrary finite-dimensional matrix used to define the trace norm, and $Q^\dagger$ is its adjoint. We use $I$ for the identity operator, with a subsystem label when needed, and $\mathrm{id}_{AB}$ for the identity channel on $AB$. Concrete operators introduced below carry their own labels.

The normalization in~\eqref{eq:negativity} is twice that of ref.~\cite{Vidal2002}. Its maximum across $A|BCD$ is one. This is a bipartite measure for the stated partition, and is not a measure of genuine four-party entanglement.

Consider
\begin{equation}
 \ket{\mathrm{CL}_4}
 =\frac{\ket{0000}+\ket{0011}+\ket{1100}-\ket{1111}}{2}.
 \label{eq:cluster}
\end{equation}
Tracing out $D$ gives
\begin{equation}
 \rho_{ABC}=\frac12\proj{\Phi^+}_{AB}\otimes\proj{0}_C
             +\frac12\proj{\Phi^-}_{AB}\otimes\proj{1}_C,
 \qquad \ket{\Phi^\pm}=\frac{\ket{00}\pm\ket{11}}{\sqrt2}.
 \label{eq:erasure}
\end{equation}
The two states of $C$ are orthogonal flags for the two Bell states. Each flagged block has negativity one, and the weights sum to one, so $\N(\rho_{ABC})=1$. A trace-preserving channel on $D$ leaves $\rho_{ABC}$ unchanged. Since discarding $D$ cannot increase negativity, the full output has negativity at least one. The upper bound for a qubit reference is also one, so the output negativity equals one. Exchanging $C,D$ establishes the same protection for a channel on $C$ alone.

This protection can also be expressed as an erasure-correction condition. The logical states $\ket{0}_B\ket{\Phi^+}_{CD}$ and $\ket{1}_B\ket{\Phi^-}_{CD}$ have identical diagonal reduced operators $I/2$ on either erased qubit and vanishing off-diagonal reduced operators. They satisfy the erasure form of the Knill--Laflamme condition~\cite{Knill1997}.

During preparation, define the Pauli flip on the fourth qubit and the phase gate on the third qubit by
\begin{equation}
 \sigma_D^x=I_A\otimes I_B\otimes I_C\otimes
 \begin{pmatrix}0&1\\1&0\end{pmatrix}_D,
 \qquad R_C(\theta)=\operatorname{diag}(1,e^{i\theta})_C.
 \label{eq:preparation}
\end{equation}
Applying $\sigma_D^x$ followed by $R_C(\theta)$ produces
\begin{equation}
 \ket{\psi_\theta}=
 \frac{\ket{00}_{AB}(\ket{01}+e^{i\theta}\ket{10})_{CD}
       +\ket{11}_{AB}(\ket{01}-e^{i\theta}\ket{10})_{CD}}{2}.
 \label{eq:state}
\end{equation}
All members have initial negativity one and the single-qubit-channel protection just proved. Their field-coupled marginal is also identical:
\begin{equation}
 \rho_{CD}(0)=\frac{\proj{01}+\proj{10}}2
             =\frac{\proj{t}+\proj{s}}2,\qquad
 \ket{t}=\frac{\ket{01}+\ket{10}}{\sqrt2},\quad
 \ket{s}=\frac{\ket{01}-\ket{10}}{\sqrt2}.
 \label{eq:marginal}
\end{equation}

The two conditional $CD$ states in~\eqref{eq:state} are normalized and orthogonal, so the reference has reduced state $I_A/2$. The flip maps $\ket{00}_{CD},\ket{11}_{CD}$ to $\ket{01}_{CD},\ket{10}_{CD}$, placing the encoding in the one-excitation subspace; $R_C(\theta)$ changes their relative phase. Together, $I_A/2$ and~\eqref{eq:marginal} show that the initial reference entanglement and detector marginal are fixed across the phase family. Their later differences therefore isolate the response of the encoded correlations.

The four-dimensional detector space splits into two orthogonal subspaces. The triplet sector is spanned by $\ket{00},\ket t,\ket{11}$; these states are symmetric under exchange of $C,D$. The singlet sector is spanned by $\ket s$, whose state vector changes sign under that exchange. A sector here means one of these subspaces, including all linear combinations of its basis states. Let $P_s=\proj{s}$ and $P_T=I-P_s$ project onto them.

A channel $\E$ describes the detector evolution after the field has been traced out. Write its Kraus representation as $\E(\rho)=\sum_\mu K_\mu\rho K_\mu^\dagger$, and impose
\begin{equation}
 [K_\mu,P_s]=0,\qquad \sum_\mu K_\mu^\dagger K_\mu=I.
 \label{eq:sectorcondition}
\end{equation}

Each Kraus operator is block diagonal in the sector decomposition. This forbids population transfer between the singlet and triplet, but allows relaxation, dephasing and mixing within the triplet. It is stronger than covariance under exchange of $C,D$. The protected object is the encoded entanglement; the detector state itself need not be stationary.

Then, for every $\theta$,
\begin{equation}
 \N\big[(\mathrm{id}_{AB}\otimes\E)(\proj{\psi_\theta})\big]\geq|\sin\theta|.
 \label{eq:sectorbound}
\end{equation}
To prove this, rewrite the initial state as
\begin{equation}
 \ket{\psi_\theta}
 =\frac{\ket{\phi_T}\ket{t}+\ket{\phi_s}\ket{s}}{\sqrt2},
 \quad
 \begin{aligned}
 \ket{\phi_T}&=\tfrac12[(1+e^{i\theta})\ket{00}+(1-e^{i\theta})\ket{11}],\\
 \ket{\phi_s}&=\tfrac12[(1-e^{i\theta})\ket{00}+(1+e^{i\theta})\ket{11}].
 \end{aligned}
 \label{eq:flags}
\end{equation}
For each normalized $AB$ state, the two amplitudes in~\eqref{eq:flags} give negativity $|1-e^{2i\theta}|/2=|\sin\theta|$. A nonselective measurement of $P_T,P_s$ after the channel leaves these states correlated with orthogonal sector flags, each of probability $1/2$. Its negativity is the probability-weighted sum $|\sin\theta|$. Monotonicity under this register-side measurement proves~\eqref{eq:sectorbound}. At $\theta=\pi/2$ modulo $\pi$, the bound saturates the maximum.

For $\theta=\pi/2$ the recovery can be coherent and deterministic. Define
\begin{equation}
 V=U_T\otimes P_T+U_s\otimes P_s,\qquad
 U_T=\operatorname{diag}(1,i)_B,\quad
 U_s=\operatorname{diag}(1,-i)_B.
 \label{eq:decoder}
\end{equation}
The decoder applies a phase gate to $B$, controlled by the sector of $CD$. At $\theta=\pi/2$, direct substitution into~\eqref{eq:flags} gives
\begin{equation}
 (I_A\otimes U_T)\ket{\phi_T}=e^{i\pi/4}\ket{\Phi^+},\qquad
 (I_A\otimes U_s)\ket{\phi_s}=e^{-i\pi/4}\ket{\Phi^+}.
 \label{eq:decoderaction}
\end{equation}
Thus both branches contain the same $AB$ Bell state, and the remaining branch phases belong to $CD$. The joint register operation obeys
\begin{equation}
 (I_A\otimes V)\ket{\psi_{\pi/2}}
 =\ket{\Phi^+}_{AB}\otimes\ket{\xi}_{CD},
 \qquad
 \ket{\xi}=\frac{e^{i\pi/4}\ket{t}+e^{-i\pi/4}\ket{s}}{\sqrt2}.
 \label{eq:decoded}
\end{equation}
Because $V$ commutes with $I_B\otimes K_\mu$, applying the same decoder after any channel satisfying~\eqref{eq:sectorcondition} yields
\begin{equation}
 (I_A\otimes V)(\mathrm{id}_{AB}\otimes\E)(\proj{\psi_{\pi/2}})
 (I_A\otimes V^\dagger)
 =\proj{\Phi^+}_{AB}\otimes\E(\proj{\xi}).
 \label{eq:recovery}
\end{equation}
The Bell pair remains recoverable even when the triplet component changes. The same decoder could also be applied before the noise, placing the Bell pair in the uncoupled $AB$ subsystem. Thus~\eqref{eq:recovery} identifies compatibility of the common noise with the encoding and recovery. The partition allows joint operations on $BCD$; restricted operations among four separated parties would require a different resource analysis. Section~\ref{sec:exact} gives an alternative readout at finite separation.

\section{Accelerated and thermal detector generators}
\label{sec:model}
We take $C,D$ to be identical two-level detectors of gap $\omega>0$ coupled to a free massless scalar field in four-dimensional Minkowski spacetime. Their trajectories have equal acceleration $a>0$ and transverse separation $L$:
\begin{align}
 x_C(\tau)&=(a^{-1}\sinh a\tau,a^{-1}\cosh a\tau,0,0),\\
 x_D(\tau)&=(a^{-1}\sinh a\tau,a^{-1}\cosh a\tau,L,0).
 \label{eq:worldlines}
\end{align}
The interaction is
\begin{equation}
 H_I(\tau)=\lambda\chi(\tau)\sum_{j=C,D}
 \big(\sigma_j^-e^{-i\omega\tau}+\sigma_j^+e^{i\omega\tau}\big)\Phi_j(\tau).
 \label{eq:interaction}
\end{equation}

Here $\lambda$ is the dimensionless weak coupling, $\chi(\tau)$ is a real switching function, and $\sigma_j^-=\ket0_j\bra1$, $\sigma_j^+=(\sigma_j^-)^\dagger$ are the detector transition operators. Equal gaps, couplings and switching functions are assumptions of exact collective protection. Free spectator evolution can be absorbed into local phases and does not change the negativity. The free detector Hamiltonian is $H_{CD}=\omega(\sigma_C^+\sigma_C^-+\sigma_D^+\sigma_D^-)$.

The field and register are initially uncorrelated. The switch-on is smooth, followed by a constant-coupling stage; the stationary generator describes that stage after bath transients. Spatially smeared $\Phi_j$ define the regulated model in section~\ref{sec:domain}. The point-profile spectral expressions quoted here are its small-profile dissipative limit.

With $\Delta\tau=\tau-\tau^\prime$ the proper-time difference, the stationary vacuum correlation along~\eqref{eq:worldlines} is
\begin{equation}
 G^+_{ij}(\Delta\tau)=-\frac{a^2}{16\pi^2}
 \left[\sinh^2\!\frac{a(\Delta\tau-i0)}2-\frac{a^2L_{ij}^2}{4}\right]^{-1},
 \qquad L_{CC}=L_{DD}=0,\quad L_{CD}=L.
 \label{eq:wightman}
\end{equation}
Let $\nu$ be the Fourier frequency, distinct from the fixed detector gap $\omega$. The $i0$ prescription selects the vacuum boundary value. Define the stationary spectrum by
\begin{equation}
 \mathcal G_{ij}(\nu)=\int_{\mathbb R}d(\Delta\tau)\,
                e^{i\nu\Delta\tau}G^+_{ij}(\Delta\tau).
 \label{eq:fourierdefinition}
\end{equation}
With this sign convention, positive $\nu$ describes de-excitation. Appendix~\ref{app:spectra} evaluates the transform and gives
\begin{align}
 \mathcal G_{\mathrm{loc}}(\nu)
 &=\frac{\nu}{2\pi(1-e^{-2\pi\nu/a})},\\
 \mathcal G_{CD}(\nu)&=\mathcal G_{\mathrm{loc}}(\nu)f_{\mathrm{acc}}(a,L,\nu),\\
 f_{\mathrm{acc}}(a,L,\omega)
 &=\frac{\sin[(2\omega/a)\arsinh(aL/2)]}
        {\omega L\sqrt{1+a^2L^2/4}} .
 \label{eq:facc}
\end{align}
Here $f_{\mathrm{acc}}(a,0,\omega)=1$. Positivity of the two-detector spectral matrix gives $|f|\leq1$, with strict inequality at $L>0$ in this continuum field model.

For the comparison, let inertial detectors at separation $L$ couple to the same massless field in a thermal state of inverse temperature $\beta=2\pi/a$. Their complete local point-profile spectra are identical, while
\begin{equation}
 f_{\mathrm{th}}(L,\omega)=\sinc(\omega L),\qquad
 \sinc x=\frac{\sin x}{x},\quad\sinc0=1.
 \label{eq:fth}
\end{equation}
The spatial distinction~\eqref{eq:facc}--\eqref{eq:fth} is established in the accelerated-atom literature~\cite{Hu2015}; below it is converted into a lifetime of encoded reference entanglement.

For either environment, the local downward and upward rates are obtained by evaluating the spectrum at the two detector transition frequencies:
\begin{equation}
 \gamma_\downarrow=\lambda^2\mathcal G_{\mathrm{loc}}(\omega),\qquad
 \gamma_\uparrow=\lambda^2\mathcal G_{\mathrm{loc}}(-\omega).
 \label{eq:rateorigins}
\end{equation}
The thermal detailed-balance relation $\mathcal G_{\mathrm{loc}}(-\omega)=e^{-2\pi\omega/a}\mathcal G_{\mathrm{loc}}(\omega)$ fixes their ratio. Substitution of~\eqref{eq:facc} gives
\begin{equation}
 q=e^{-2\pi\omega/a},\qquad
 \gamma_\downarrow=\frac{\lambda^2\omega}{2\pi(1-q)},\qquad
 \gamma_\uparrow=q\gamma_\downarrow,\qquad
 \Gamma=\gamma_\downarrow+\gamma_\uparrow,\quad z=\Gamma\tau.
 \label{eq:rates}
\end{equation}

Here $q$ is the upward-to-downward rate ratio, $\Gamma$ sets the local relaxation scale, and $z$ is dimensionless time. The symbol $f$ below denotes the spatial factor of the selected environment evaluated at $\omega$.

The off-diagonal field spectrum couples the two detectors. Diagonalizing the spectral matrix identifies the corresponding collective channels. At each frequency,
\begin{equation}
 [\mathcal G_{ij}(\nu)]_{i,j=C,D}
 =\mathcal G_{\mathrm{loc}}(\nu)
   \begin{pmatrix}1&f(\nu)\\f(\nu)&1\end{pmatrix},
 \qquad v_\pm=\frac{(1,\pm1)^{\mathsf T}}{\sqrt2}.
 \label{eq:spectralmatrix}
\end{equation}
Here $v_\pm$ are normalized column vectors of the $C,D$ coefficients, and $\mathsf T$ denotes transpose. Equal or opposite components give the symmetric or antisymmetric combination.
The eigenvalues are $\mathcal G_{\mathrm{loc}}(\nu)[1\pm f(\nu)]$. For both environments $f(\nu)$ is real and even, so excitation and de-excitation use the same collective modes. The Kubo--Martin--Schwinger (KMS) thermal condition fixes their rate ratio to $q$. The secular approximation averages terms oscillating at distinct transition frequencies. Together with weak coupling, it gives a Gorini--Kossakowski--Sudarshan--Lindblad (GKSL) generator~\cite{Gorini1976,Lindblad1976}.

The same $\lambda,\omega,a$ give the same $\Gamma,q$ for the point-profile comparison. For regulated profiles, the local rates may instead be matched by calibrating the coupling strengths. The secular weak-coupling generator retains the collective terms because the two gaps are degenerate:
\begin{equation}
 \dot\rho=-i[H_{\mathrm{eff}},\rho]
 +\sum_{\eta=\pm1}(1+\eta f)
 \big(\gamma_\downarrow\D[J_\eta]\rho+
      \gamma_\uparrow\D[J_\eta^\dagger]\rho\big),
 \quad J_\pm=\frac{\sigma_C^-\pm\sigma_D^-}{\sqrt2},
 \label{eq:gksl}
\end{equation}
Here $\rho$ is the $CD$ density matrix and the dot means $d/d\tau$. The dissipator is $\D[J]\rho=J\rho J^\dagger-\{J^\dagger J,\rho\}/2$, where braces denote the anticommutator. Its first term transfers population into the state reached by the jump $J$; the other terms remove the corresponding population. We denote the full generator on the right-hand side by $\mathscr L$. The resulting four-qubit channel acts as the identity on $A,B$.

The renormalized Hamiltonian contribution $H_{\mathrm{eff}}$ in the interaction-picture generator is diagonal in the Dicke basis
\begin{equation}
 \ket{g}=\ket{00},\quad\ket{t},\quad
 \ket{e}=\ket{11},\quad\ket{s}.
 \label{eq:dicke}
\end{equation}

This basis resolves symmetric and antisymmetric transitions, reflecting the collective-radiation structure of Dicke states~\cite{Dicke1954}. The nonzero downward actions are
\begin{equation}
 J_+\ket e=\ket t,\quad J_+\ket t=\ket g,
 \qquad J_-\ket e=\ket s,\quad J_-\ket s=-\ket g.
 \label{eq:ladders}
\end{equation}
At $f=1$ the antisymmetric rate vanishes and the singlet does not participate in any transition. For $f<1$ that ladder transfers population between sectors. The minus sign in~\eqref{eq:ladders} affects an amplitude, while population rates involve its squared modulus.

For example, the exchange term is $\Omega_{\mathrm{ex}}(\sigma_C^+\sigma_D^-+\sigma_D^+\sigma_C^-)$. It changes the phase of the $t$--$s$ coherence but not the populations or the negativity derived below. Thus it need not be discarded. A Hamiltonian mixing Dicke states requires a different population reduction. Smearing is kept finite when defining or renormalizing the Lamb shift; its unregulated coincident point limit is not used.

\section{Exact negativity and the phase optimum}
\label{sec:exact}
Let $\E_z$ be the completely positive trace-preserving channel generated on $CD$ at proper time $\tau=z/\Gamma$. To evaluate the reference--register negativity, first write the initial state as
\begin{equation}
 \ket{\psi_\theta}=\frac{\ket{00}\ket{\zeta_+}+\ket{11}\ket{\zeta_-}}{\sqrt2},
 \qquad
 \ket{\zeta_+}=\frac{\ket{01}+e^{i\theta}\ket{10}}{\sqrt2},\quad
 \ket{\zeta_-}=\frac{\ket{01}-e^{i\theta}\ket{10}}{\sqrt2}.
 \label{eq:uv}
\end{equation}
The two terms share the same register evolution. Define their cross operator by
\begin{equation}
 \mathsf W_\theta(z)=\E_z(\ket{\zeta_+}\bra{\zeta_-}).
 \label{eq:crossoperator}
\end{equation}
Partial transpose on $A$ moves this operator into an independent block with eigenvalues $\pm\varsigma_j(\mathsf W_\theta)/2$, where $\varsigma_j$ are singular values. The remaining blocks are positive and have total trace one. Subtracting this contribution according to~\eqref{eq:negativity} gives
\begin{equation}
 \N_\theta(z)=\|\mathsf W_\theta(z)\|_1=\|\E_z(\ket{\zeta_+}\bra{\zeta_-})\|_1.
 \label{eq:traceidentity}
\end{equation}
Appendix~\ref{app:trace} shows the block construction. Thus the entanglement calculation reduces to the evolution of one operator on the four-dimensional detector space.

The Dicke basis separates the diagonal and off-diagonal parts of that operator. Define $P_t=\proj t$, $P_s=\proj s$ and the phase coefficients
\begin{equation}
 \alpha_\theta=\frac{\sin\theta}{2},\quad
 b_\theta=\frac{1+\cos\theta}{2},\quad
 c_\theta=\frac{1-\cos\theta}{2}.
 \label{eq:abc}
\end{equation}
Substituting $\ket{01}=(\ket t+\ket s)/\sqrt2$ and $\ket{10}=(\ket t-\ket s)/\sqrt2$ gives
\begin{equation}
 \ket{\zeta_+}\bra{\zeta_-}=i\alpha_\theta(P_t-P_s)
       +b_\theta\ket{t}\bra{s}+c_\theta\ket{s}\bra{t}.
 \label{eq:operator}
\end{equation}
We therefore need the evolved population difference $P_t-P_s$ and the two coherences $\ket t\bra s,\ket s\bra t$.

The population equations follow directly from~\eqref{eq:gksl}. Define $p_j=\bra j\rho_{CD}\ket j$ for $j=g,t,e,s$. Since $H_{\mathrm{eff}}$ is diagonal in this basis, its commutator has zero diagonal entries. Each dissipative jump contributes a gain at its destination and a loss at its origin. For example, $g\to t$ and $e\to t$ increase $p_t$, whereas $t\to g$ and $t\to e$ decrease it. Equation~\eqref{eq:ladders} therefore gives
\begin{equation}
 \frac{dp_t}{d\tau}
 =(1+f)\gamma_\uparrow p_g+(1+f)\gamma_\downarrow p_e
  -(1+f)(\gamma_\downarrow+\gamma_\uparrow)p_t.
 \label{eq:tripletbalance}
\end{equation}
Dividing by $\Gamma$, because $z=\Gamma\tau$, introduces the dimensionless downward and upward rates
\begin{equation}
 r_\pm=\frac{(1\pm f)\gamma_\downarrow}{\Gamma}
       =\frac{1\pm f}{1+q},\qquad
 \frac{(1\pm f)\gamma_\uparrow}{\Gamma}=qr_\pm.
 \label{eq:scaledrates}
\end{equation}
The same gain-minus-loss calculation for all four Dicke states yields
\begin{align}
 \frac{dp_g}{dz}&=-q(r_++r_-)p_g+r_+p_t+r_-p_s,\nonumber\\
 \frac{dp_t}{dz}&=qr_+p_g-(1+q)r_+p_t+r_+p_e,\nonumber\\
 \frac{dp_e}{dz}&=qr_+p_t-(r_++r_-)p_e+qr_-p_s,\nonumber\\
 \frac{dp_s}{dz}&=qr_-p_g+r_-p_e-(1+q)r_-p_s.
 \label{eq:populationrows}
\end{align}
These equations close on the populations even when the initial density matrix has coherences. Collecting their coefficients in the order $g,t,e,s$ gives
\begin{equation}
 \frac{d\boldsymbol p}{dz}=M\boldsymbol p,\qquad
 \boldsymbol p=(p_g,p_t,p_e,p_s)^{\mathsf T}.
 \label{eq:populationevolution}
\end{equation}
The coefficient matrix is
\begin{equation}
 M=
 \begin{pmatrix}
 -q(r_++r_-)&r_+&0&r_-\\
 qr_+&-(1+q)r_+&r_+&0\\
 0&qr_+&-(r_++r_-)&qr_-\\
 qr_-&0&r_-&-(1+q)r_-
 \end{pmatrix}.
 \label{eq:M}
\end{equation}
The second row of $M$ is the equation for $p_t$ and the fourth is the equation for $p_s$. Every column sums to zero, so $\sum_jp_j$ is conserved. The off-diagonal entries are nonnegative transition rates.

Now compare evolution from the initial triplet state $P_t$ and from the initial singlet state $P_s$. Define $p_j^{(t)}=\bra j\E_z(P_t)\ket j$, $p_j^{(s)}=\bra j\E_z(P_s)\ket j$ and $d_j=p_j^{(t)}-p_j^{(s)}$. Both distributions satisfy~\eqref{eq:populationevolution}, so their difference satisfies $d\boldsymbol d/dz=M\boldsymbol d$. The initial difference is $(0,1,0,-1)^{\mathsf T}$, and hence
\begin{equation}
 \boldsymbol d(z)=e^{M z}(0,1,0,-1)^{\mathsf T},\qquad
 \boldsymbol d=(d_g,d_t,d_e,d_s)^{\mathsf T},
 \label{eq:d}
\end{equation}
Each distribution has total probability one, implying $\sum_jd_j=0$. The entries of $\boldsymbol d$ can be negative because they are differences of probabilities.

The off-diagonal part has a simpler evolution. For $\ket t\bra s$, each jump-gain term vanishes because a given collective jump annihilates either $\ket t$ or $\ket s$. The two downward anticommutators together contribute $-\gamma_\downarrow\ket t\bra s$: their weights are $(1+f)/2$ and $(1-f)/2$. The upward terms similarly contribute $-\gamma_\uparrow\ket t\bra s$. If $E_t,E_s$ are the corresponding eigenenergies of $H_{\mathrm{eff}}$, the full result is
\begin{equation}
 \mathscr L(\ket t\bra s)
 =-[\Gamma+i(E_t-E_s)]\ket t\bra s.
 \label{eq:coherencegenerator}
\end{equation}
Integrating this scalar equation and using $z=\Gamma\tau$ gives
\begin{equation}
 \E_z(\ket t\bra s)=e^{-z-i\varphi(z)}\ket t\bra s,
 \qquad \varphi(z)=\frac{E_t-E_s}{\Gamma}z,
 \label{eq:coherencelaw}
\end{equation}
The reverse coherence has the complex-conjugate phase. The decay modulus is $e^{-z}$ for both coherences, independently of $f$.

Combining the population difference and coherence evolution in~\eqref{eq:operator} now gives, in the ordered Dicke basis,
\begin{equation}
 \mathsf W_\theta(z)=
 \begin{pmatrix}
 i\alpha_\theta d_g&0&0&0\\
 0&i\alpha_\theta d_t&0&b_\theta e^{-z-i\varphi}\\
 0&0&i\alpha_\theta d_e&0\\
 0&c_\theta e^{-z+i\varphi}&0&i\alpha_\theta d_s
 \end{pmatrix}.
 \label{eq:crossmatrix}
\end{equation}
Here $\varphi$ abbreviates $\varphi(z)$. The scalar entries contribute $|\alpha_\theta|(|d_g|+|d_e|)$ to the trace norm. Let $Y$ be the block on the $t,s$ rows and columns. If its singular values are $\varsigma_1,\varsigma_2$, then
\begin{equation}
 \|Y\|_1^2=(\varsigma_1+\varsigma_2)^2
          =\tr(YY^\dagger)+2|\det Y|.
 \label{eq:blocknorm}
\end{equation}
The identity follows from $\varsigma_1^2+\varsigma_2^2=\tr(YY^\dagger)$ and $\varsigma_1\varsigma_2=|\det Y|$. Since $b_\theta c_\theta=\alpha_\theta^2$, the required terms are
\begin{align}
 \tr(YY^\dagger)
 &=\alpha_\theta^2(d_t^2+d_s^2)
   +(b_\theta^2+c_\theta^2)e^{-2z},\nonumber\\
 \det Y&=-\alpha_\theta^2(d_td_s+e^{-2z}).
 \label{eq:blockingredients}
\end{align}
The Hamiltonian phases cancel in the determinant. Adding the scalar contributions to $\|Y\|_1$ establishes the exact negativity
\begin{align}
 \N_\theta(z)
 ={}&|\alpha_\theta|(|d_g|+|d_e|)\nonumber\\
 &+\left[
 \alpha_\theta^2(d_t^2+d_s^2)+(b_\theta^2+c_\theta^2)e^{-2z}
 +2\alpha_\theta^2|d_td_s+e^{-2z}|
 \right]^{1/2}.
 \label{eq:exact}
\end{align}
The determinant modulus remains inside the formula, so no assumption about the sign of $d_td_s+e^{-2z}$ is needed.

The phase dependence can now be isolated. Put $h_\theta=|\sin\theta|$ and $E_z=e^{-2z}$. Using $b_\theta^2+c_\theta^2=1-h_\theta^2/2$, define
\begin{equation}
 K_z=\frac{d_t^2+d_s^2-2E_z+2|d_td_s+E_z|}{4}.
 \label{eq:K}
\end{equation}
The triangle inequality gives $|d_td_s+E_z|\geq E_z-|d_td_s|$. Consequently $4K_z\geq(|d_t|-|d_s|)^2\geq0$. Equation~\eqref{eq:exact} becomes
\begin{equation}
 \N_\theta(z)=\frac{h_\theta}{2}(|d_g|+|d_e|)
                  +\sqrt{E_z+h_\theta^2K_z}
 \label{eq:phase}
\end{equation}
At any finite $z$, $E_z>0$, and differentiation with respect to $h_\theta\in[0,1]$ gives
\begin{equation}
 \frac{\partial\N_\theta}{\partial h_\theta}
 =\frac{|d_g|+|d_e|}{2}
  +\frac{h_\theta K_z}{\sqrt{E_z+h_\theta^2K_z}}
 \geq0.
 \label{eq:phasemonotonicity}
\end{equation}
The negativity is therefore largest when $h_\theta=1$, namely at $\theta=\pi/2$ modulo $\pi$. Its smallest value occurs at $h_\theta=0$:
\begin{equation}
 e^{-z}=\N_0(z)\leq\N_\theta(z)\leq\N_{\pi/2}(z).
 \label{eq:optimum}
\end{equation}
This optimum holds at finite separation as well as at exact collective coupling. Every phase has nonzero negativity at finite time within this generator. At $f=0$, the formula reduces to $e^{-z}$ for every phase.

To obtain a two-qubit resource, measure $CD$ in the Dicke basis and apply a conditional phase gate to $B$. The detector--field coupling is ended before this readout. For outcome $j$, the unnormalized conditional state is $\rho_{AB}^{(j)}=\bra j\rho_{ABCD}(z)\ket j$. Its off-diagonal entry follows from~\eqref{eq:crossmatrix}:
\begin{equation}
 \bra{00}\rho_{AB}^{(j)}\ket{11}
 =\frac12\bra j\mathsf W_\theta(z)\ket j
 =\frac{i\alpha_\theta d_j(z)}2.
 \label{eq:conditionalcoherence}
\end{equation}
A phase gate on $B$ makes this entry real and nonnegative. Its angle is fixed by the known preparation and dynamics; a zero entry requires no correction. Retaining every outcome and then discarding the measurement record adds the magnitudes of these conditional coherences. The resulting negativity is
\begin{equation}
 \N_{\mathrm{rec},\theta}(z)
 =\frac{|\sin\theta|}{2}\sum_{j=g,t,e,s}|d_j(z)|
 =|\sin\theta|\,\mathcal T_z\leq\N_\theta(z),
 \label{eq:recoverfinite}
\end{equation}
Here $\mathcal T_z=\frac12\sum_j|p_j^{(t)}-p_j^{(s)}|$ is the total-variation distance between the two population distributions. This distance lies between zero and one and measures how well the two initial sectors can still be distinguished by the detector measurement. The upper bound in~\eqref{eq:recoverfinite} follows because measurement, correction and discarding $CD$ are all operations on the register side of the partition.

The summed diagonal entries on $AB$ remain $1/2,1/2$. Thus the recovered state is
\begin{equation}
 \rho_{AB}^{\mathrm{rec}}=
 \frac{\proj{00}+\proj{11}}2
 +\frac{\N_{\mathrm{rec},\theta}}2
   (\ket{00}\bra{11}+\ket{11}\bra{00}).
 \label{eq:recoveredstate}
\end{equation}
Partial transpose on $A$ gives eigenvalues $1/2,1/2,\N_{\mathrm{rec},\theta}/2,-\N_{\mathrm{rec},\theta}/2$. Equation~\eqref{eq:negativity} then reproduces~\eqref{eq:recoverfinite}. This is a deterministic readout with no postselection. It gives an operational lower bound at every time, and the next section shows that this bound agrees with the full negativity to leading order on the slow scale.

The recovered state is a mixture of $\ket{\Phi^+},\ket{\Phi^-}$ with weights $(1+\N_{\mathrm{rec},\theta})/2$ and $(1-\N_{\mathrm{rec},\theta})/2$. Its maximal Bell-state overlap and the average fidelity of standard two-qubit teleportation are~\cite{Horodecki1999}
\begin{equation}
 \mathcal F_{\mathrm{Bell}}=\frac{1+\N_{\mathrm{rec},\theta}}2,
 \qquad \overline{\mathcal F}_{\mathrm{tel}}
 =\frac{2\mathcal F_{\mathrm{Bell}}+1}{3}
 =\frac{2+\N_{\mathrm{rec},\theta}}3.
 \label{eq:teleportation}
\end{equation}
The average is uniform over unknown pure qubit inputs. Whenever $\N_{\mathrm{rec},\theta}>0$, the fidelity exceeds the classical threshold $2/3$. This relation concerns the recovered state and the specified protocol.

\section{Metastability and the matched-temperature comparison}
\label{sec:slow}
At $f=1$ the antisymmetric ladder is absent, so the singlet stays fixed. The triplet ladder obeys detailed balance: the upward-to-downward ratio $q$ gives equilibrium populations in the ratio $1:q:q^2$ for $g,t,e$. Normalizing these populations gives
\begin{equation}
 \rho_T^{\mathrm{eq}}
 =\frac{\proj{g}+q\proj{t}+q^2\proj{e}}{Z},
 \qquad Z=1+q+q^2.
 \label{eq:tripleteq}
\end{equation}
Consequently
\begin{equation}
 \lim_{z\to\infty}\N_\theta(z)=|\sin\theta|,
 \qquad \N_{\pi/2}(z)=1,\quad \N_0(z)=e^{-z}.
 \label{eq:collective}
\end{equation}
The first statement is the equilibrated version of the sector bound; the second holds at every time by the decoder construction. A phase changes the fate of reference entanglement while the marginal $CD$ evolution remains identical.

For $|f|<1$ and $q>0$, both upward and downward ladders are active. Their operators span the local lowering and raising operators on $C,D$, so the channel relaxes to the unique full-rank state
\begin{equation}
 \rho_{CD}^{\mathrm{eq}}=
 \frac{(\proj{0}+q\proj{1})_C\otimes
       (\proj{0}+q\proj{1})_D}{(1+q)^2}.
 \label{eq:thermaleq}
\end{equation}
All traceless operators decay, and~\eqref{eq:traceidentity} gives $\N_\theta(\infty)=0$. The $q=0$ boundary also relaxes to $\proj{g}$ at $|f|<1$, although its stationary state is not full rank. In fact, direct solution of~\eqref{eq:M} gives the useful exact boundary formula
\begin{equation}
 \left.\N_\theta(z)\right|_{q=0}
 =e^{-z}\left[h_\theta|\sinh(fz)|+
             \sqrt{1+h_\theta^2\sinh^2(fz)}\right],
 \quad h_\theta=|\sin\theta|.
 \label{eq:zerotemp}
\end{equation}
In particular $\N_{\pi/2}(z)=e^{-(1-|f|)z}$ there. At $|f|<1$, zero temperature therefore does not restore permanent entanglement at distinct detector positions.

Set $\epsilon=1-f\ll1$ at fixed $0<q<1$. The symmetric rate $r_+=(2-\epsilon)/(1+q)$ remains of order one, while the antisymmetric rate $r_-=\epsilon/(1+q)$ is small. The triplet therefore relaxes before substantial transfer to or from the singlet occurs. Define its total population by $p_T=p_g+p_t+p_e$.

After the fast transient, the triplet populations are $p_g=p_T/Z+O(\epsilon)$, $p_t=qp_T/Z+O(\epsilon)$ and $p_e=q^2p_T/Z+O(\epsilon)$. The last row of~\eqref{eq:populationrows}, converted back to proper time, then gives
\begin{align}
 \frac{dp_s}{d\tau}
 &=\epsilon(\gamma_\uparrow p_g+\gamma_\downarrow p_e-\Gamma p_s)\nonumber\\
 &=\epsilon\Gamma
   \left[\frac{q+q^2}{(1+q)Z}p_T-p_s\right]
   +O(\epsilon^2\Gamma)\nonumber\\
 &=\epsilon\Gamma\left(\frac qZp_T-p_s\right)
   +O(\epsilon^2\Gamma).
 \label{eq:slowflux}
\end{align}
The simplification uses $(q+q^2)/(1+q)=q$. Thus the effective triplet-to-singlet rate is $\epsilon\Gamma q/Z$, and the singlet-to-triplet rate is $\epsilon\Gamma$. Conservation of $p_T+p_s=1$ fixes the opposite flux into the triplet. Together the two equations are
\begin{equation}
 \frac{d}{d\tau}\begin{pmatrix}p_T\\p_s\end{pmatrix}
 =\epsilon\Gamma
 \begin{pmatrix}-q/Z&1\\q/Z&-1\end{pmatrix}
 \begin{pmatrix}p_T\\p_s\end{pmatrix}
 +O(\epsilon^2\Gamma).
 \label{eq:sectorrates}
\end{equation}
The leading two-by-two matrix has zero determinant, reflecting probability conservation. Its trace is $-(1+q/Z)$, so this is its nonzero eigenvalue. Restoring the factor $\epsilon\Gamma$ gives the slow decay rate
\begin{equation}
 \kappa=\Gamma\epsilon \mathcal C(q)+O(\Gamma\epsilon^2),
 \qquad \mathcal C(q)=1+\frac qZ=\frac{(1+q)^2}{1+q+q^2}.
 \label{eq:kappa}
\end{equation}
The rate determines the decay of a sector population difference. To obtain the entanglement amplitude, also project the initial difference onto the fast-equilibrated triplet and singlet. Appendix~\ref{app:slow} derives this projection and controls the finite-dimensional limit. For fixed slow time $u=\epsilon z>0$, it gives
\begin{equation}
 \boldsymbol d(u/\epsilon)\longrightarrow
 e^{-\mathcal C(q)u}
 \left[\frac{(1,q,q^2,0)^{\mathsf T}}{Z}
                  -(0,0,0,1)^{\mathsf T}\right].
 \label{eq:slowdifference}
\end{equation}
The $t$--$s$ coherence is proportional to $e^{-u/\epsilon}$ and vanishes on this scale. The absolute components of the vector in brackets sum to $Z/Z+1=2$. Substitution into~\eqref{eq:exact} and~\eqref{eq:recoverfinite} therefore yields
\begin{equation}
 \lim_{\epsilon\to0}\N_\theta(u/\epsilon)
 =\lim_{\epsilon\to0}\N_{\mathrm{rec},\theta}(u/\epsilon)
 =|\sin\theta|e^{-\mathcal C(q)u}.
 \label{eq:slowlimit}
\end{equation}
This limit separates two stages. Fast relaxation at $z=O(1)$ first reduces the general-phase negativity from one to a value near $|\sin\theta|$. The later leakage occurs at $u=O(1)$, or $z=O(\epsilon^{-1})$. In the intermediate window $1\ll z\ll\epsilon^{-1}$, the negativity consequently remains near $|\sin\theta|$. Convergence is on compact intervals with $u>0$; the slow formula does not describe the initial fast transient.

\begin{figure}[t]
 \centering
 \begin{subfigure}[t]{0.48\textwidth}
  \centering\includegraphics[width=\linewidth]{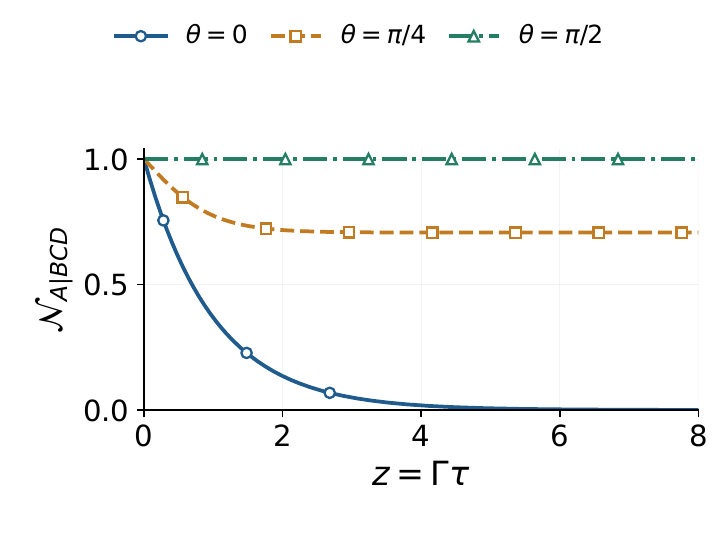}
  \caption{}\label{fig:collective}
 \end{subfigure}\hfill
 \begin{subfigure}[t]{0.48\textwidth}
  \centering\includegraphics[width=\linewidth]{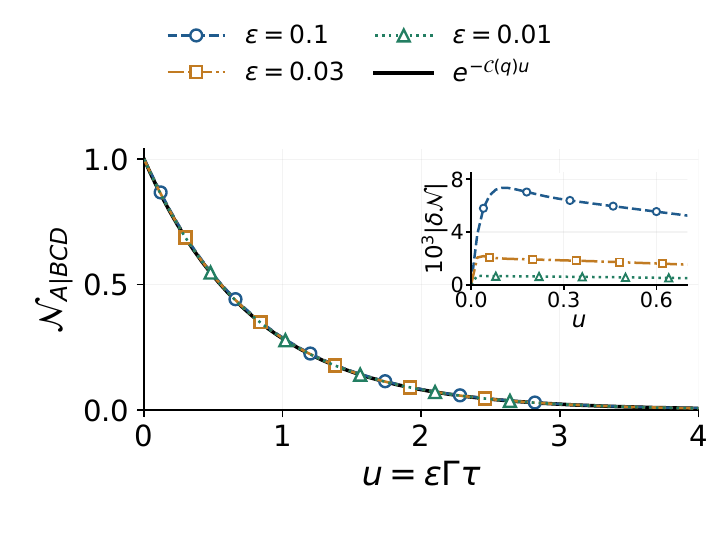}
  \caption{}\label{fig:slowdecay}
 \end{subfigure}
 \caption{Phase selection and the slow limit, with $q=0.4$. (a) Exact negativity at $f=1$ for three locally equivalent initial states; line patterns and markers distinguish the phases. (b) Marked curves are the exact results at $\theta=\pi/2$ and $\epsilon=0.1,0.03,0.01$, plotted against $u=\epsilon\Gamma\tau$. The black solid line is $e^{-\mathcal C(q)u}$. The inset shows $10^3|\delta\N|$, with $\delta\N=\N_{\pi/2}-e^{-\mathcal C(q)u}$ at the same $u$. Markers sample the exact curves without displacing their data. The results follow equations~\eqref{eq:exact} and~\eqref{eq:slowlimit}.}
 \label{fig:protection}
\end{figure}

At the optimal phase the slow amplitude is one. Exact collective coupling then has no entanglement loss, while a nonzero separation starts the loss immediately. Expanding~\eqref{eq:exact} at small $z$ gives
\begin{equation}
 \N_{\pi/2}(z)=1-\epsilon z+O(z^2),\qquad f\geq0.
 \label{eq:initialslope}
\end{equation}
Figure~\ref{fig:protection} separates the effect of preparation phase from this weak leakage. Panel (a) fixes $f=1$. Panel (b) plots the finite-$\epsilon$ results against $u=\epsilon\Gamma\tau$. Curves at different $\epsilon$ therefore correspond to different physical time scales, and their overlap tests the common slow law. The inset resolves their differences from that law. The choice $q=0.4$ corresponds to $a/\omega=-2\pi/\ln0.4\simeq6.86$ and makes the thermal-population contribution to $\mathcal C(q)$ visible.

To express this lifetime in terms of detector geometry, expand the spatial factors in~\eqref{eq:facc} and~\eqref{eq:fth}. Appendix~\ref{app:spectra} gives the intermediate Taylor terms. At small transverse separation, the result is
\begin{align}
 \epsilon_{\mathrm{acc}}&=\frac{\omega^2+a^2}{6}L^2+O(L^4),&
 \epsilon_{\mathrm{th}}&=\frac{\omega^2}{6}L^2+O(L^4).
 \label{eq:epsgeometry}
\end{align}
Hence the lifetime $\tau_{\mathrm{life}}=\kappa^{-1}$ has leading accelerated value
\begin{equation}
 \tau_{\mathrm{life}}^{\mathrm{acc}}
 \simeq\frac{6Z}{\Gamma(1+q)^2(\omega^2+a^2)L^2}.
 \label{eq:lifetime}
\end{equation}
At the same local $\Gamma,q$, the ratio of slow rates is
\begin{equation}
 \lim_{L\to0}\frac{\kappa_{\mathrm{acc}}}{\kappa_{\mathrm{th}}}
 =R,\qquad R=1+\frac{a^2}{\omega^2}.
 \label{eq:rateratio}
\end{equation}
The factor $\mathcal C(q)$ and the coupling strength cancel. For example, $a/\omega=1$ and $2$ give ratios $2$ and $5$. The spatial spectra differ even though a single point-profile detector has the same local spectrum in both environments. The protected-phase lifetime isolates that spatial difference.

For a more direct time-domain comparison introduce
$w=\Gamma\omega^2L^2\tau/6$. At fixed $a/\omega>0$, the point-profile small-separation limit yields
\begin{equation}
 \N_{\pi/2}^{\mathrm{th}}\longrightarrow e^{-\mathcal C(q)w},
 \qquad
 \N_{\pi/2}^{\mathrm{acc}}\longrightarrow e^{-\mathcal C(q)Rw}.
 \label{eq:contrastcurves}
\end{equation}
For the limiting curves, define the contrast $\Delta\N^{(0)}(w)=e^{-\mathcal C(q)w}-e^{-\mathcal C(q)Rw}$. Since $R>1$, it is positive for $w>0$ and vanishes at both endpoints $w=0$ and $w\to\infty$. Its stationary point satisfies
\begin{equation}
 \begin{aligned}
 \frac{d\Delta\N^{(0)}}{dw}
 &=\mathcal C(q)
   [R e^{-\mathcal C(q)Rw}-e^{-\mathcal C(q)w}]=0\\
 &\Longleftrightarrow\quad e^{\mathcal C(q)(R-1)w}=R.
 \end{aligned}
 \label{eq:contraststationary}
\end{equation}
There is a unique positive solution. The derivative is positive before it and negative after it, so this point gives the global maximum:

\begin{equation}
 w_*=\frac{\ln R}{\mathcal C(q)(R-1)},\qquad
 \Delta\N_{\max}^{(0)}
 =(1-R^{-1})R^{-1/(R-1)}.
 \label{eq:optimalcontrast}
\end{equation}

Substitution of $w_*$ into the two exponentials gives the second expression in~\eqref{eq:optimalcontrast}. At $a/\omega=1$ the limiting contrast is $1/4$; at $a/\omega=2$ it is approximately $0.5350$. The contrast is an entanglement observable, not a minimum discrimination error probability. The fixed decoding measurement~\eqref{eq:recoverfinite} gives the same leading contrast in the recovered $AB$ negativity.

The corresponding physical observation time is
\begin{equation}
 \tau_*\simeq\frac{6\ln R}
 {\Gamma\omega^2L^2\mathcal C(q)(R-1)}.
 \label{eq:optimaltime}
\end{equation}
At finite separation the optimal time is obtained from~\eqref{eq:exact}; equation~\eqref{eq:optimaltime} is its small-separation leading value at fixed $a/\omega>0$. The recovered standard-teleportation fidelity has the same leading optimum time and maximum contrast
\begin{equation}
 \Delta\overline{\mathcal F}_{\mathrm{tel},\max}^{(0)}
 =\frac13(1-R^{-1})R^{-1/(R-1)}.
 \label{eq:fidelitycontrast}
\end{equation}
This follows from the linear dependence on recovered negativity in~\eqref{eq:teleportation}. The leading fidelity contrasts at $a/\omega=1$ and $2$ are $1/12$ and approximately $0.1783$, respectively.

\begin{figure}[t]
 \centering
 \begin{subfigure}[t]{0.48\textwidth}
  \centering\includegraphics[width=\linewidth]{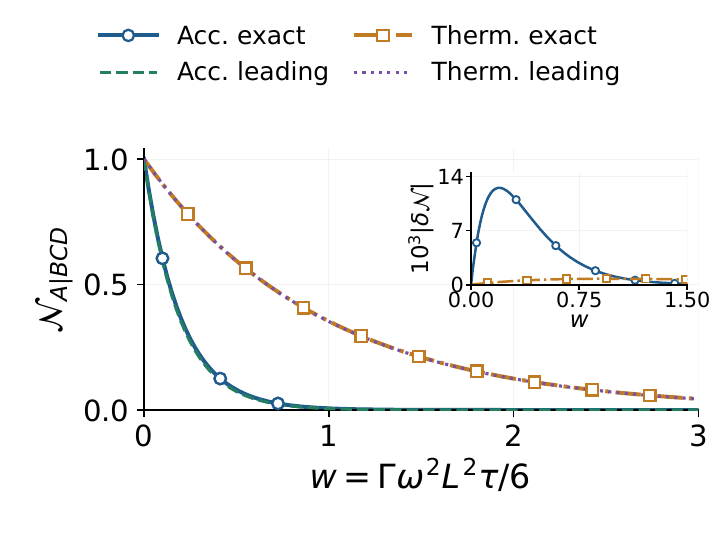}
  \caption{}\label{fig:comparison}
 \end{subfigure}\hfill
 \begin{subfigure}[t]{0.48\textwidth}
  \centering\includegraphics[width=\linewidth]{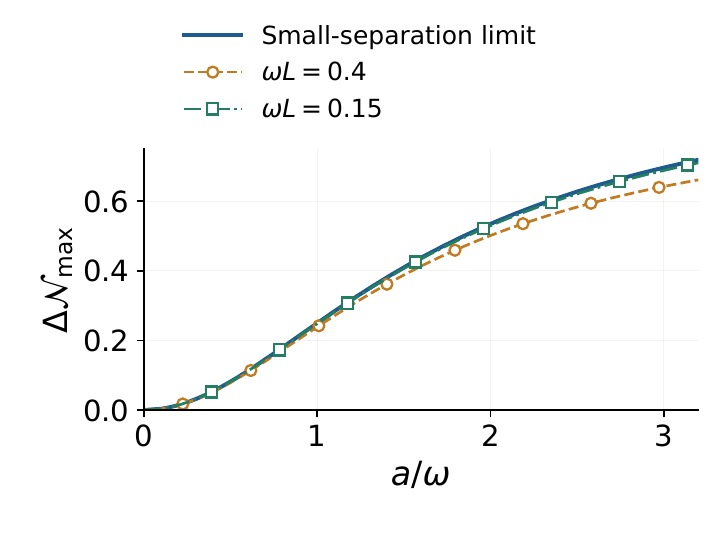}
  \caption{}\label{fig:contrast}
 \end{subfigure}
 \caption{Entanglement contrast at matched local temperature and relaxation rate. (a) Protected-phase results at $a/\omega=2$, $\omega L=0.2$, plotted against $w=\Gamma\omega^2L^2\tau/6$. The solid curve with circles and the dash-dotted curve with squares are the exact accelerated and thermal results. Dashed and dotted curves are their respective leading limits~\eqref{eq:contrastcurves}. The inset gives $10^3|\delta\N|$, where $\delta\N=\N^{\mathrm{exact}}-\N^{\mathrm{leading}}$, with the colors of the exact curves. (b) The unmarked solid curve is~\eqref{eq:optimalcontrast}. Marked curves maximize the exact negativity difference at the indicated finite separations; their line patterns and symbols match the legend.}
 \label{fig:thermal}
\end{figure}

Figure~\ref{fig:thermal} presents complementary aspects of the matched comparison: panel (a) shows the resources at the same observation time, while panel (b) gives the maximum available contrast as a function of $a/\omega$ and checks it against finite separations. The matching fixes local rates and temperature, rather than the cross spectrum.

The two time limits are distinct. For every finite $L>0$, eventual relaxation erases the reference--register entanglement, whereas an exactly common coupling preserves the Bell pair:
\begin{equation}
 \lim_{L\to0^+}\lim_{\tau\to\infty}\N_{\pi/2}=0,\qquad
 \lim_{\tau\to\infty}\N_{\pi/2}\big|_{L=0}=1.
 \label{eq:noncommute}
\end{equation}
These statements concern the effective generator; the exactly common regulated coupling additionally has the non-Markovian protection described next.

Independent evolution of the full four-qubit density matrix agrees with~\eqref{eq:exact}, and sector-preserving channel tests agree with the recovery identity~\eqref{eq:recovery}. The numerical slow decay rate reproduces the small-$\epsilon$ expansion~\eqref{eq:kappa}.

\section{Common coupling and the lifetime approximation}
\label{sec:domain}
Spatial smearing models a detector that samples the field over a finite region with a prescribed weight. Let $\boldsymbol\xi=(\xi_x,\xi_y,\xi_z)$ denote displacement from its centre, and let the smooth real profile $F_r(\boldsymbol\xi)$ have support radius $r<a^{-1}$, even transverse dependence and normalization $\int d^3\xi\,F_r=1$. For the scalar field operator $\phi(x)$, the weighted field operators are
\begin{equation}
 \begin{gathered}
 \Phi_j(\tau)=\int d^3\xi\,F_r(\boldsymbol\xi)
 \phi\big((a^{-1}+\xi_x)\sinh a\tau,
          (a^{-1}+\xi_x)\cosh a\tau,
          y_j+\xi_y,\xi_z\big),\\
 y_C=0,\qquad y_D=L.
 \end{gathered}
 \label{eq:smearing}
\end{equation}
Here $r$ is the detector radius, $L$ is the distance between the detector centres, and $\tau$ is the central trajectory's proper time. The two profiles differ only by a transverse translation.

The profiles remain inside the right Rindler wedge and are stationary under the boost evolution, so the vacuum correlations obey KMS at inverse temperature $2\pi/a$. The finite profile is retained when defining the Lamb shift.

At $L=0$ the two weighted field operators coincide. Identical gaps, couplings and switching functions then give
\begin{equation}
 [H_I(\tau),P_s]=[H_{CD},P_s]=0.
 \label{eq:microsym}
\end{equation}
The full detector--field unitary therefore commutes with the singlet projector $P_s$. Its bath matrix elements give Kraus operators with the same property. The sector bound~\eqref{eq:sectorbound} and fixed recovery~\eqref{eq:recovery} consequently hold for this common coupling without a Markov approximation, for any initially independent bath state.

At finite detector size, write the spatial factor as $f_r(L)$ and retain the notation $\epsilon=1-f_r(L)$. For identical profiles, $f_r(0)=1$; reflection symmetry and a finite fourth spectral moment give $\epsilon=c_rL^2+O(L^4)$, where $c_r$ depends on the profile and environment. Appendix~\ref{app:stability} derives this coefficient and states the conditions for its point-profile limit. The slow law~\eqref{eq:slowlimit} applies with this spatial factor. Matching the local rates leaves the ratio $c_r^{\mathrm{acc}}/c_r^{\mathrm{th}}$, whose small-profile value is~\eqref{eq:rateratio}.

For $L>0$, the lifetime is calculated from the weak-coupling master equation~\eqref{eq:gksl}. It requires $\Gamma\ll\omega$ and $\Gamma\tau_B\ll1$, where $\tau_B$ is the time over which bath correlations decay. The renormalized Hamiltonian scales must also be compatible with these approximations~\cite{Kaplanek2020,Han2025}. The equal detector gaps allow the collective cross terms to survive secular averaging.

The lifetime grows as the separation decreases, so microscopic accuracy must be assessed on that growing time scale. At fixed $L>0$ and $r>0$, assume weak-coupling convergence uniformly on bounded intervals of $z=\Gamma_{\lambda,r}\tau$, where $\Gamma_{\lambda,r}$ is the regulated local relaxation rate~\cite{Davies1974}.

First take $\lambda\to0$ at this fixed geometry. At fixed slow time $u>0$, the corresponding $z=u/[1-f_r(L)]$ is finite, so the convergence assumption applies. Then take $L\to0$. Equation~\eqref{eq:slowlimit} gives
\begin{equation}
 \lim_{L\to0^+}\lim_{\lambda\to0}
 \N_\theta^{\mathrm{mic}}\!
 \left(\frac{u}{\Gamma_{\lambda,r}[1-f_r(L)]}\right)
 =|\sin\theta|e^{-\mathcal C(q)u}.
 \label{eq:iterated}
\end{equation}
This ordered limit supplies the microscopic interpretation of the lifetime. A simultaneous limit $L=L(\lambda)$ would require an additional uniform error estimate.

Additional noise can be assessed directly within the effective dynamics. Let $\mathscr L$ and $\widetilde{\mathscr L}$ be the time-independent generators before and after the perturbation, both producing completely positive trace-preserving evolution. Define the dimensionless perturbation size $\delta$ by
\begin{equation}
 \|\widetilde{\mathscr L}-\mathscr L\|_\diamond\leq\delta\Gamma .
 \label{eq:perturbation}
\end{equation}
The diamond norm measures the action of the map while allowing a reference system. Appendix~\ref{app:stability} bounds the accumulated change in negativity:
\begin{equation}
 |\widetilde\N_\theta(\tau)-\N_\theta(\tau)|
 \leq2\delta\Gamma\tau
 =\frac{2\delta u}{\epsilon}
 \quad\left(\tau=\frac{u}{\epsilon\Gamma}\right).
 \label{eq:error}
\end{equation}
For fixed $u$, the error vanishes if $\delta/\epsilon\to0$. Thus additional generator terms must be small compared with the slow rate $\epsilon\Gamma$. In a near-degenerate description retaining collective terms, a sufficient condition for a gap mismatch $\Delta\omega$ is $|\Delta\omega|\ll\epsilon\Gamma$; an additional dephasing rate must satisfy the same scale comparison.

\section{Conclusions}
\label{sec:conclusion}
A local preparation phase controls the recoverable entanglement of the four-qubit encoding without changing its initial reference entanglement or detector marginal. At $\theta=\pi/2$, sector-preserving noise admits a fixed Bell-pair decoder. At zero separation, identical couplings and profiles enforce this protection by symmetry. At finite separation, the detector generator establishes the same phase as optimal at every time.

Small separation opens the antisymmetric channel and sets the slow resource lifetime. Matching the local response of accelerated vacuum and an inertial thermal bath leaves the leading rate ratio $1+a^2/\omega^2$. The derived optimum time maximizes their entanglement contrast, which is also accessible to leading order in the recovered two-qubit resource. The lifetime is a weak-coupling prediction whose microscopic interpretation uses the ordered limits in section~\ref{sec:domain}.

\appendix
\section{Spectral factors and their small-separation expansion}
\label{app:spectra}
Put $b_L=\arsinh(aL/2)$, $\ell_L=2b_L/a$ and $\beta=2\pi/a$. To transform~\eqref{eq:wightman}, continue the proper-time difference $\Delta\tau$ into the complex plane. At $L>0$, zeros of the denominator occur when $a(\Delta\tau-i0)/2=\pm b_L+in\pi$, giving the poles
\begin{equation}
 \Delta\tau=\pm\ell_L+i0+in\beta,\qquad n\in\mathbb Z.
\end{equation}
The $i0$ prescription is a positive infinitesimal displacement fixed by the vacuum correlation. Writing the complex time difference as $x+iy$ with real $x,y$, the Fourier factor at $\nu>0$ obeys $|e^{i\nu(x+iy)}|=e^{-\nu y}$, so the contour closes in the upper half-plane and encloses the poles with $n\geq0$.

Each pole is simple. Its residue is the numerator of the integrand divided by the derivative of the denominator at that pole. In the lowest strip this derivative is
\begin{equation}
 \begin{aligned}
 \left.\frac{d}{d\Delta\tau}
       \sinh^2\frac{a(\Delta\tau-i0)}2\right|_{\Delta\tau=\pm\ell_L+i0}
 &=\left.\frac a2\sinh[a(\Delta\tau-i0)]\right|_{\Delta\tau=\pm\ell_L+i0}\\
 &=\pm\frac{a^2L\cosh b_L}{2}.
 \end{aligned}
 \label{eq:residuederivative}
\end{equation}
Multiplication by the prefactor $-a^2/(16\pi^2)$ in~\eqref{eq:wightman} and the Fourier factor gives the residue sum
\begin{equation}
 -\frac{e^{i\nu\ell_L}}{8\pi^2L\cosh b_L}
 +\frac{e^{-i\nu\ell_L}}{8\pi^2L\cosh b_L}
 =-\frac{i\sin(\nu\ell_L)}{4\pi^2L\cosh b_L}.
\end{equation}
A shift by $in\beta$ multiplies this sum by $e^{-n\beta\nu}$. Summing $\sum_{n=0}^\infty e^{-n\beta\nu}=(1-e^{-\beta\nu})^{-1}$ and applying the contour factor $2\pi i$ therefore yields
\begin{equation}
 \mathcal G_{CD}(\nu)
 =\frac{\sin(\nu\ell_L)}
 {2\pi L\cosh b_L(1-e^{-\beta\nu})}.
 \label{eq:residues}
\end{equation}
Since $\ell_L/L\to1$ and $\cosh b_L\to1$ as $L\to0$, this expression gives the local spectrum in~\eqref{eq:facc}. Negative frequency follows from thermal detailed balance, $\mathcal G_{ij}(-\nu)=e^{-\beta\nu}\mathcal G_{ij}(\nu)$ for $\nu>0$.

For inertial detectors in a thermal state, start from the momentum representation
\begin{equation}
 G_\beta^+(\Delta\tau,\mathbf L)=
 \int\frac{d^3k}{2(2\pi)^3k}
 \big[(1+n_\beta(k))e^{-ik\Delta\tau+i\mathbf k\cdot\mathbf L}
       +n_\beta(k)e^{ik\Delta\tau-i\mathbf k\cdot\mathbf L}\big],
 \qquad n_\beta(k)=\frac1{e^{\beta k}-1}.
\end{equation}
Here $k=|\mathbf k|$ is the field momentum magnitude, $\mathbf L$ is the separation vector with length $L$, and $n_\beta(k)$ is the thermal occupation number. At positive frequency, the first time-dependent term gives
\begin{equation}
 \int_{\mathbb R}d\Delta\tau\,e^{i(\nu-k)\Delta\tau}
 =2\pi\delta(\nu-k).
 \label{eq:thermaltransform}
\end{equation}
The other term gives $\delta(\nu+k)$ and contributes nothing for $\nu,k>0$. The radial momentum integral is therefore evaluated at $k=\nu$. Choosing the polar axis along $\mathbf L$, with polar angle $\vartheta_{\mathbf k}$ and solid-angle element $d\Omega_{\mathbf k}$, gives
\begin{equation}
 \begin{aligned}
 \left.\int d\Omega_{\mathbf k}\,e^{i\mathbf k\cdot\mathbf L}\right|_{k=\nu}
 &=2\pi\int_0^\pi d\vartheta_{\mathbf k}\,
       \sin\vartheta_{\mathbf k}\,e^{i\nu L\cos\vartheta_{\mathbf k}}\\
 &=4\pi\sinc(\nu L).
 \end{aligned}
 \label{eq:thermalangular}
\end{equation}
Combining the radial and angular factors produces
\begin{equation*}
 \mathcal G_{CD}^{\mathrm{th}}(\nu)
 =\frac{\nu[1+n_\beta(\nu)]}{2\pi}\sinc(\nu L)
 =\frac{\nu\sinc(\nu L)}{2\pi(1-e^{-\beta\nu})}.
\end{equation*}
At $L=0$, $\sinc(\nu L)=1$, so the complete local spectrum agrees with the accelerated one. Division by that local spectrum gives~\eqref{eq:fth}.

The small-separation coefficients follow by expanding the accelerated numerator and denominator separately. First,
\begin{equation}
 \frac2a\arsinh(aL/2)=L-\frac{a^2L^3}{24}+O(L^5),\qquad
 \sqrt{1+a^2L^2/4}=1+\frac{a^2L^2}{8}+O(L^4).
\end{equation}
Inserting the first expansion into $\sin(\omega\ell_L)$, and using $\sin x=x-x^3/6+O(x^5)$, gives
\begin{equation}
 \sin(\omega\ell_L)
 =\omega L\left[1-\left(\frac{\omega^2}{6}
                            +\frac{a^2}{24}\right)L^2+O(L^4)\right].
 \label{eq:smallseparationnumerator}
\end{equation}
The inverse square-root factor is $1-a^2L^2/8+O(L^4)$. Their product in~\eqref{eq:facc} is
\begin{equation}
 \begin{aligned}
 f_{\mathrm{acc}}
 &=\left[1-\left(\frac{\omega^2}{6}+\frac{a^2}{24}\right)L^2\right]
  \left[1-\frac{a^2L^2}{8}\right]+O(L^4)\\
 &=1-\frac{\omega^2+a^2}{6}L^2+O(L^4).
 \end{aligned}
 \label{eq:expandedfactor}
\end{equation}
For the thermal factor, $\sinc(\omega L)=1-\omega^2L^2/6+O(L^4)$. Subtracting the two factors from one gives~\eqref{eq:epsgeometry}.

\section{Trace-norm reduction}
\label{app:trace}
Applying the channel to~\eqref{eq:uv} gives the full state
\begin{align}
 \rho_{ABCD}(z)=\frac12\big[&
 \proj{00}_{AB}\otimes\E_z(\proj{\zeta_+})
 +\proj{11}_{AB}\otimes\E_z(\proj{\zeta_-})\nonumber\\
 &+\ket{00}\bra{11}_{AB}\otimes\mathsf W_\theta
 +\ket{11}\bra{00}_{AB}\otimes\mathsf W_\theta^\dagger\big].
 \label{eq:channeloutput}
\end{align}
Partial transpose acts on $A$ alone, so $(\ket{00}\bra{11})^{T_A}=\ket{10}\bra{01}$. In the ordered $AB$ basis $00,01,10,11$, the cross terms therefore form an independent block on the $01,10$ rows and columns:
\begin{equation}
 \frac12\begin{pmatrix}0&\mathsf W_\theta^\dagger\\\mathsf W_\theta&0\end{pmatrix},
 \qquad \mathsf W_\theta(z)=\E_z(\ket{\zeta_+}\bra{\zeta_-}).
\end{equation}
A singular-value decomposition of $\mathsf W_\theta$ reduces this block to pairs of eigenvalues $\pm\varsigma_j(\mathsf W_\theta)/2$. The two diagonal blocks on $00,11$ are positive, each with trace $1/2$, because $\E_z$ is positive and trace preserving. Thus the full trace norm is $1+\sum_j\varsigma_j(\mathsf W_\theta)$, proving~\eqref{eq:traceidentity}.

In the Dicke basis, the scalar entries of $\mathsf W_\theta$ contribute $|\alpha_\theta d_g|+|\alpha_\theta d_e|$. Its remaining block is
\begin{equation}
 Y=\begin{pmatrix}
 i\alpha_\theta d_t&b_\theta e^{-z-i\varphi}\\
 c_\theta e^{-z+i\varphi}&i\alpha_\theta d_s
 \end{pmatrix}.
\end{equation}
The squared singular values of $Y$ are the eigenvalues of $YY^\dagger$. Their sum is $\tr(YY^\dagger)$, and their product is $\det(YY^\dagger)=|\det Y|^2$. Taking the nonnegative square root of the product gives $\varsigma_1\varsigma_2=|\det Y|$. Expanding $(\varsigma_1+\varsigma_2)^2$ then proves~\eqref{eq:blocknorm}. Direct evaluation gives~\eqref{eq:blockingredients}; the factors $e^{\pm i\varphi}$ cancel in the off-diagonal product. Adding the scalar contributions establishes~\eqref{eq:exact}.

The zero-temperature boundary can be obtained without spectral perturbation. At $q=0$, upward jumps are absent. The initial difference has $d_e(0)=0$, so $d_e$ remains zero; the triplet and singlet components decay independently, and trace conservation fixes $d_g$. The solution is
\begin{equation}
 d_t=e^{-(1+f)z},\quad d_s=-e^{-(1-f)z},\quad
 d_e=0,\quad d_g=e^{-(1-f)z}-e^{-(1+f)z}.
\end{equation}
Substitution into~\eqref{eq:exact} gives~\eqref{eq:zerotemp}. The absolute value in that expression also covers negative $f$.

\section{Slow-sector projection}
\label{app:slow}
Insert $f=1-\epsilon$ into~\eqref{eq:M}. Since $r_+=(2-\epsilon)/(1+q)$ and $r_-=\epsilon/(1+q)$, the matrix is exactly $M=M_0+\epsilon M_1$, with $M_0=M|_{f=1}$ and
\begin{equation}
 M_1=\frac1{1+q}
 \begin{pmatrix}
 0&-1&0&1\\
 -q&1+q&-1&0\\
 0&-q&0&q\\
 q&0&1&-(1+q)
 \end{pmatrix}.
 \label{eq:populationperturbation}
\end{equation}
At $\epsilon=0$, the triplet equilibrium and the stationary singlet give two right zero-eigenvectors:
\begin{equation}
 \boldsymbol r_T=(1,q,q^2,0)^{\mathsf T}/Z,\qquad
 \boldsymbol r_s=(0,0,0,1)^{\mathsf T}.
\end{equation}
The corresponding left vectors are $\boldsymbol\ell_T=(1,1,1,0)$ and $\boldsymbol\ell_s=(0,0,0,1)$. They extract total triplet and singlet populations and satisfy $\boldsymbol\ell_i\boldsymbol r_j=\delta_{ij}$. The remaining two population eigenvalues are
\begin{equation}
 -2+\frac{2\sqrt q}{1+q},\qquad
 -2-\frac{2\sqrt q}{1+q}.
\end{equation}
For fixed $q>0$, both are bounded away from zero. They describe the fast relaxation within the triplet, while the two-dimensional zero-eigenspace is the subspace in which slow leakage must be resolved.

Projecting $M_1$ with the left and right vectors gives the effective sector generator. Its singlet-row entries are
\begin{equation}
 \boldsymbol\ell_sM_1\boldsymbol r_T=\frac qZ,\qquad
 \boldsymbol\ell_sM_1\boldsymbol r_s=-1.
 \label{eq:projectionentries}
\end{equation}
The triplet-row entries have opposite signs, because $M_1$ preserves total probability. Hence
\begin{equation}
 [\boldsymbol\ell_iM_1\boldsymbol r_j]_{i,j=T,s}
 =\begin{pmatrix}-q/Z&1\\q/Z&-1\end{pmatrix}.
\end{equation}
This matrix has eigenvalues zero and $-\mathcal C(q)$, with difference eigenvector $(1,-1)^{\mathsf T}$. The initial vector $(0,1,0,-1)^{\mathsf T}$ has projected sector components $(1,-1)^{\mathsf T}$. Its slow projection is therefore $\boldsymbol r_T-\boldsymbol r_s$, with no total stationary component.

Finite-dimensional perturbation theory separates this slow subspace from the fast modes. At $z=u/\epsilon$, the fast contributions vanish and the difference mode evolves as
\begin{equation}
 e^{M u/\epsilon}(0,1,0,-1)^{\mathsf T}
 \longrightarrow e^{-\mathcal C(q)u}(\boldsymbol r_T-\boldsymbol r_s)
 \quad(\epsilon\to0,\ u>0).
\end{equation}
The coherence also vanishes as $e^{-u/\epsilon}$. The two distributions $\boldsymbol r_T,\boldsymbol r_s$ have disjoint supports and each sums to one, so $\sum_j|(\boldsymbol r_T-\boldsymbol r_s)_j|=2$. This gives~\eqref{eq:slowlimit} from either~\eqref{eq:exact} or~\eqref{eq:recoverfinite}. At fixed $q>0$, the error is $O(\epsilon)$ on compact positive intervals of $u$, together with an exponentially small fast contribution. Appendix~\ref{app:trace} gives the zero-temperature boundary independently.

The initial slope follows from a different, short-time expansion. The first-order terms in $e^{Mz}$ give $d_t=1-(1+f)z+O(z^2)$, $d_s=-1+(1-f)z+O(z^2)$, $d_g=2fz/(1+q)+O(z^2)$ and $d_e=2qfz/(1+q)+O(z^2)$. For $f\geq0$, substitution into~\eqref{eq:exact} gives~\eqref{eq:initialslope}.

\section{Spatial smearing and slow-time stability}
\label{app:stability}
At fixed profile radius $r$, a transverse translation by $L$ contributes the phase $e^{ik_yL}$ to the cross spectrum, with $k_y$ the transverse momentum. Reflection symmetry removes its imaginary part. The normalized spatial factor is therefore an average of $\cos(k_yL)$ against a positive normalized spectral measure $d\mu_r$. Expanding the cosine, with a finite fourth moment, gives
\begin{equation}
 f_r(L)=\int\cos(k_yL)\,d\mu_r
       =1-c_rL^2+O(L^4),\qquad
 c_r=\frac12\int k_y^2\,d\mu_r.
 \label{eq:regulatedgeometry}
\end{equation}
This establishes the quadratic leakage used in section~\ref{sec:domain}. For a smooth approximate-identity family of profiles whose on-shell moments at the gap $\omega$ converge, the coefficients tend to $c_r^{\mathrm{acc}}\to(\omega^2+a^2)/6$ and $c_r^{\mathrm{th}}\to\omega^2/6$ as $r\to0$. Their ratio then gives~\eqref{eq:rateratio} after matching the local rates. The point-profile limit is taken in these dissipative spectral coefficients, while the microscopic interaction and Lamb shift are defined at finite $r$.

Next consider two time-independent GKSL generators. Duhamel's identity gives
\begin{equation}
 e^{\tau\widetilde{\mathscr L}}-e^{\tau\mathscr L}
 =\int_0^\tau dv\,e^{(\tau-v)\widetilde{\mathscr L}}
       (\widetilde{\mathscr L}-\mathscr L)e^{v\mathscr L}.
\end{equation}
The diamond norm of each trace-preserving completely positive channel is one. Submultiplicativity therefore bounds the integrand by $\delta\Gamma$, and integration over $v\in[0,\tau]$ gives
$\|e^{\tau\widetilde{\mathscr L}}-e^{\tau\mathscr L}\|_\diamond\leq\delta\Gamma\tau$. Applying the two channels to the same normalized initial state consequently bounds the trace norm of the output-state difference by $\delta\Gamma\tau$.
For any two four-qubit states $\rho,\widetilde\rho$,
\begin{equation}
 |\N(\widetilde\rho)-\N(\rho)|
 \leq\|(\widetilde\rho-\rho)^{T_A}\|_1
 \leq2\|\widetilde\rho-\rho\|_1.
\end{equation}
The factor two is the induced trace-norm bound for partial transpose on the qubit reference. Combining these inequalities proves~\eqref{eq:error}.

\end{document}